\documentclass[11pt,a4paper]{article}
\usepackage{jheppub}
\usepackage{amsmath}
\usepackage[most]{tcolorbox}
\usepackage{dsfont}
\usepackage{ulem}
\usepackage{natbib}
\usepackage{xcolor}
\usepackage[hang,flushmargin]{footmisc}
\usepackage{tikz-cd}
\usepackage{enumitem}
\usepackage{amsfonts,amssymb, amscd,amsmath,latexsym,amsbsy,bm}
\usepackage{stmaryrd}
\usepackage{todonotes}

\makeatletter\renewcommand{\@biblabel}[1]{#1.}\makeatother

\newtcolorbox{empheqboxed}{colback=gray!20, 
	colframe=white,
	width=\textwidth,
	sharpish corners,
	top=0mm, 
	bottom=0pt
}

\title{Lens Hyperbolic Modular Double}

\author{Yağmur Bülbül$^{a}$, Ilmar Gahramanov$^{a,b}$, Ali Mert Yetkin$^{a}$, Reyhan Yumuşak$^{a}$}
\affiliation{$^a$ Department of Physics, Bogazici University,\\ 34342 Bebek, Istanbul, Türkiye\\[-0.5cm]
	
$^b$ Center for Mathematics and its Apllications, Khazar University, \\ Mehseti St. 41, AZ1096, Baku, Azerbaijan}

\emailAdd{bulbulygmr@gmail.com, ilmar.gahramanov@bogazici.edu.tr, yknalimert@gmail.com, reyhanyumusakk@gmail.com }

\abstract{We construct the lens hyperbolic modular double, a new algebraic structure whose intertwining operator produces a lens hyperbolic hypergeometric solution of the Yang--Baxter equation.}

\keywords{modular double, star-triangle relation, integrable lattice spin model, Ising-type model, Yang-Baxter equation, gauge/YBE correspondence.}

\begin{document}
	\maketitle
	\flushbottom

\section{Introduction}

\section*{Introduction}

Faddeev’s modular double~\cite{Faddeev:1999fe} is an elegant algebraic construction that unifies two mutually dual quantum groups. 
The deformation parameters \( q = e^{i\pi b^2} \) and \( \tilde q = e^{i\pi b^{-2}} \) are related by the modular transformation \( b \leftrightarrow b^{-1} \), reflecting the intrinsic self-duality of the underlying structure. 
This duality lies at the heart of various developments in mathematical physics, where the modular double plays a central role in the study of integrable systems, conformal field theory, and quantum geometry.

In this work, we consider an integrable two-dimensional lattice model of statistical mechanics~\cite{Gahramanov:2016ilb}, which generalizes the Spiridonov model\footnote{
The Spiridonov model itself is a generalization of the Faddeev--Volkov model~\cite{Bazhanov:2007vg}.
}~\cite{Spiridonov:2010em}. 
In this model, the spin variables take both continuous and discrete values, and the Boltzmann weights are expressed in terms of hyperbolic gamma functions. 
The associated star--triangle relation (the Yang--Baxter equation) for this model was obtained via the gauge/YBE correspondence, specifically from three-dimensional \(\mathcal{N}=2\) supersymmetric dual theories defined on the lens space.

The main goal of the present work is to investigate the algebraic structure underlying the Ising-like model introduced in~\cite{Gahramanov:2016ilb} and its relation to quantum integrable systems. 
We demonstrate that the hyperbolic modular double introduced in~\cite{Chicherin:2015mfv} admits a natural extension, which we refer to as the lens hyperbolic modular double. 
In the limit \( r \to 1 \), this algebra reduces to the standard hyperbolic modular double. 
Using the known star--triangle relation for the Ising-like lattice spin model, we construct a solution to the Yang--Baxter equation that exhibits the symmetry of the lens hyperbolic modular double. 
This solution is realized as an integral operator acting on a pair of infinite-dimensional representation spaces of the algebra.

\section{A Brief Introduction to the Sklyanin Algebra}

As is well known, establishing the integrability of a model reduces to proving the Yang–Baxter equation
\begin{equation}
    \mathbb{R}_{12}(u-v)\mathbb{R}_{13}(u)\mathbb{R}_{23}(v)= \mathbb{R}_{23}(v)\mathbb{R}_{13}(u)\mathbb{R}_{12}(u-v) \;. \label{eq:YBE}
\end{equation}
The operators $\mathbb{R}_{ik}$ act on $\mathbb{V}_1 \otimes \mathbb{V}_2 \otimes \mathbb{V}_3$ and provide a general solution. In what follows, we introduce an algebraic structure that emerges when one of the spaces, $\mathbb{V}_1$, is arbitrary, while the remaining two spaces, $\mathbb{V}_2$ and $\mathbb{V}_3 = \mathbb{C}^2$, are two-dimensional. In this case, the $R$-matrix can be reduced to what is known as a Lax operator \cite{Sklyanin:1982tf}
\begin{equation}
    \mathbb{R}_{13}(u):=\textnormal{L}_{13}(u)= \sum_{a=0}^3w_a(u)\sigma_a \otimes \mathbf{S}^a= \begin{pmatrix}
        w_0(u)\mathbf{S}^0+w_3(u)\mathbf{S}^3 & w_1(u)\mathbf{S}^1-iw_2(u)\mathbf{S}^2\\
        w_1(u) \mathbf{S}^1+iw_2(u)\mathbf{S}^2 & w_0(u)\mathbf{S}^0-w_3(u)\mathbf{S}^3
    \end{pmatrix} \;,
\end{equation}
where $w_a(u)=\frac{\theta_{a+1}(u+\eta)}{\theta_{a+1}(\eta)}$ and $\mathbf{S}^a$ are arbitrary operators. This allows the reduction of the Yang--Baxter equation (\ref{eq:YBE}) as well\footnote{A standard example of $\mathbb{R}_{12}$ is the $R$-matrix obtained by Baxter for the eight-vertex model \cite{Baxter:1982zz}.}
\begin{equation}
    \mathbb{R}_{12}(u-v)\textnormal{L}_{13}(u)\textnormal{L}_{23}(v)=\textnormal{L}_{23}(v)\textnormal{L}_{13}(u)\mathbb{R}_{12}(u-v) \;. \label{eq:reducedYBE}
\end{equation}
In \cite{Sklyanin:1982tf}, Sklyanin showed that for the operators $\mathbf{S}^a$, following relations are equivalent to the relation (\ref{eq:reducedYBE}), remarkably forming an algebra
\begin{subequations}
    \begin{equation}
     \mathbf{S}^{\alpha}\mathbf{S}^{\beta}-\mathbf{S}^{\beta}\mathbf{S}^{\alpha}=i(\mathbf{S}^0\mathbf{S}^{\gamma}+\mathbf{S}^{\gamma}\mathbf{S}^0) \;, \label{SAcom1}
    \end{equation}
    \begin{equation}
        \mathbf{S}^0\mathbf{S}^{\alpha}-\mathbf{S}^{\alpha}\mathbf{S}^0=i \mathbf{J}_{\beta \gamma}(\mathbf{S}^{\beta}\mathbf{S}^{\gamma}+\mathbf{S}^{\gamma}\mathbf{S}^{\beta}) \;. \label{SAcom2}
    \end{equation}
\end{subequations} 
The structure constants $\mathbf{J}_{\beta\gamma}$ in equation (\ref{SAcom2}) can be written as
\begin{align}
        J_1 = \frac{\theta_2(2\gamma)\theta_2(0)}{(\theta_2(\gamma))^2} \;,  &&
        J_2= \frac{\theta_3(2\gamma)\theta_3(0)}{(\theta_3(\gamma))^2} \;, &&
        J_3= \frac{\theta_4(2\gamma)\theta_4(0)}{(\theta_4(\gamma))^2} \;,
\end{align}
where the definition for $\theta_i(z)$ can be found in (\ref{appendix1}).
A trigonometric degeneration of the Sklyanin algebra can be obtained in the limit $h \to 0$ where  \( h = e^{i\pi\tau} \) is the elliptic nome. This was shown by Zabrodin and Gorsky in \cite{Gorsky:1993mh} via a redefinition of the generators $\mathbf{S}^a$
\begin{subequations}
\begin{align}
A &= ih^{-1/4}\frac{(S^0-\tanh{(\pi \gamma)}S^3)}{4\sinh{(\pi\gamma)}} \;, 
&\quad 
D &= ih^{-1/4}\frac{(S^0+\tanh{(\pi \gamma)}S^3)}{4\sinh{(\pi\gamma)}} \;, \\[1em]
C &= ih^{\pm 1/4} \frac{(S^1-S^2)}{2\sinh{(2\pi\gamma)}} \;,
&\quad 
B &= ih^{-3/4}\frac{(S^1+iS^2)}{8\sinh{(2\pi\gamma)}} \;.
\end{align}
\end{subequations}

To proceed, one can substitute these into the commutators (\ref{SAcom1}) and (\ref{SAcom2}). This substitution is made possible by the realization of the generators $\mathbf{S}^{a}$ as difference operators \cite{Sklyanin:1982tf}. Denoting $q=\exp{(-2\pi\gamma)}$, the limit $h \to 0$ then admits the following commutators

\begin{subequations}
\begin{align}
DC &= qCD \;, &\qquad CA &= qAC \;, \\[0.8em]
[A,D] &= \frac{1}{4}(q-q^{-1})^3C^2 \;, 
&\qquad [B,C] &= \frac{(A^2-D^2)}{q-q^{-1}} \;,
\end{align}
\begin{align}
AB - qBA &= qDB - BD 
= -\frac{1}{4}(q^2-q^{-2})(DC-CA) \;.
\end{align}
\label{trigosklyanin}
\end{subequations}

A common procedure to investigate the Sklyanin algebra representations is breaking down of the $R$-matrix and finding an operator $M$ that acts as an intertwining operator for the generators of the algebra, which amounts to the following
\begin{subequations}
    \begin{align}
        \textnormal{M}(g)A(g)=A(-g)\textnormal{M}(g)\;, && \textnormal{M}(g)B(g)=B(-g)\textnormal{M}(g)\;,
    \end{align}
    \begin{align}
        \textnormal{M}(g)C(g)=C(-g)\textnormal{M}(g) \;,&&\textnormal{M}(g)D(g)=D(-g)\textnormal{M}(g) \;.
    \end{align}
\end{subequations}
For more details we refer the reader to the works \cite{Sklyanin:1982tf,Derkachov:2012iv,Chicherin:2014dya,spiridonov2009continuousbiorthogonalityelliptichypergeometric,Antonov:1997zc,Chicherin:2015mfv} and references therein.

\section{Lens Hyperbolic Solution to the Star-Triangle Equation}

Here we consider an integrable lattice model of statistical mechanics introduced in \cite{Gahramanov:2016ilb}. The related models were also studied in \cite{Eren:2019ibl,Bozkurt:2020gyy,Catak:2021coz,Gahramanov:2022jxz,Catak:2024ygo,Mullahasanoglu:2023nes}. In this integrable model similar to Ising-type models, the spins are placed on the square lattice. In this case, however, each spin $\sigma_i$ carries both a continuous $x_i$ and a discrete component $m_i$, 
\begin{equation}
\sigma_i = (x_i, m_i) \;.
\end{equation}
The Boltzmann weight $W_{\alpha_i,\tilde{\alpha}}(\sigma_i,\sigma_j)$ (or $W_{\alpha_i,\tilde{\alpha}}(x_i,x_j;m_i,m_j)$) is assigned to a horizontal edge with spins $\sigma_i$ and $\sigma_j$ that is crossed by rapidity lines $\alpha_i$ and $\tilde{\alpha_i}$. Similarly, the Boltzmann weight $\overline{W}_{\alpha_i,\tilde{\alpha}}(\sigma_i,\sigma_j)$ (or $\overline{W}_{\alpha_i,\tilde{\alpha}}(x_i,x_j;m_i,m_j)$) is assigned to a vertical edge with spins $\sigma_i$ and $\sigma_j$ that is crossed by rapidity lines $\alpha_i$ and $\tilde{\alpha_i}$. The edge Boltzmann weights are given by a product of eight hyperbolic gamma functions, defined as follows
\begin{subequations}
\begin{equation}
    \begin{aligned}
       W_{\alpha_i,\tilde{\alpha}_i}(x_i,x_j;m_i,m_j)
       =\gamma_h(-\alpha_i+x_i\pm x_j,-\tilde{\alpha}_i+m_i\pm m_j)\\ \times \gamma_h(-\alpha_i-x_i\pm x_j,-\tilde{\alpha_i}-m_i\pm m_j) \;,
    \end{aligned}
\end{equation}
\begin{equation}
    \begin{aligned}
       \overline{W}_{\alpha_i,\tilde{\alpha}_i}(x_i,x_j;m_i,m_j)
       =\gamma_h(\eta+\alpha_i+x_i\pm x_j,n_\eta +\tilde{\alpha}_i+m_i\pm m_j)\\ \times \gamma_h(\eta+\alpha_i-x_i\pm x_j,n_\eta +\tilde{\alpha_i}-m_i\pm m_j) \;,
    \end{aligned}
\end{equation}
\end{subequations}
and the self interaction contributes as
\begin{equation}
\begin{aligned}
 S(\sigma_0) = \frac{ \epsilon (n)}{\gamma_h(\pm 2u, \pm 2n)}\:,
\end{aligned}
\end{equation}
where \( \sigma_0=(u,n)\), $\eta=\sum_i\alpha_i$, $n_\eta=\sum_i\tilde{\alpha}_i$, and $\epsilon (n) $ is defined as $\epsilon(0)=\epsilon(\lfloor \frac{r}{2}\rfloor)=1$ and $\epsilon(n)=2$ otherwise. 
Here $\gamma_h(u,y)$ denotes the multiplication of two hyperbolic gamma functions, defined as 
\begin{equation}
    \gamma_h(u,y) = \gamma^{(2)}(u+\omega_1y;\omega_1r,\omega_1+\omega_2)\:\:\gamma^{(2)}(u+\omega_2(r-y);\omega_2r,\omega_1+\omega_2),
\end{equation}
and $\gamma^{(2)}(z;\omega_1,\omega_2)$ is the hyperbolic gamma function defined as
\begin{subequations}
    \begin{equation}
        \gamma^{(2)}(u;\omega_1,\omega_2):= e^{B_{2,2}(u;\omega_1,\omega_2)} \frac{(e^{2\pi i  u /\omega_1}\Tilde{q};\Tilde{q})_\infty}{(e^{2\pi i  u /\omega_2};q)_\infty} \;,
    \end{equation}
    \begin{equation}
        (t;q)_\infty=\prod_{k=0}^\infty(1-tq^k) \;,
    \end{equation}
\end{subequations}
where $q=e^{2\pi i \omega_1 / \omega_2}$, $ \Tilde{q}=e^{-2\pi i \omega_2/\omega_1} $ and
$
    B_{2,2}(u;\omega_1,\omega_2)=\frac{1}{\omega_1\omega_2} \left( \left( u -\frac{\omega_1+\omega_2}{2}\right)^2-\frac{\omega_1^2+\omega_2^2}{12}\right)
$.

The Boltzmann weights are defined in terms of hyperbolic gamma functions have the following property \cite{Spiridonov:2010em,Faddeev:1999fe}
\begin{equation}
    W_{\alpha_i,\tilde{\alpha_i}}(\sigma_i,\sigma_j) = \overline{W}_{\alpha_i,\tilde{\alpha_i}}(\sigma_i,\sigma_j)\;.
\end{equation}
As is well known, for models of this type it is sufficient to verify the star–triangle relation in order to establish the integrability of the theory. The star–triangle equation can be written in the following form
\begin{align}
\begin{aligned}
   \sum_{m_0} \int dx_0 \: S(\sigma_0) \: W_{\alpha_1,\Tilde{\alpha}_1}(\sigma_1,\sigma_0)W_{\alpha_2,\Tilde{\alpha}_2}(\sigma_2,\sigma_0)W_{\alpha_3,\Tilde{\alpha}_3}(\sigma_3,\sigma_0) 
    \makebox[10em]{}
   \\ \makebox[6em]{}
   =\mathcal{R} \: W_{\eta-\alpha_1,n_\eta-\Tilde{\alpha}_1}(\sigma_1,\sigma_2)W_{\eta-\alpha_2,n_\eta-\Tilde{\alpha}_2}(\sigma_1,\sigma_3)W_{\eta-\alpha_3,n_\eta-\Tilde{\alpha}_3}(\sigma_2,\sigma_3) 
   \label{str}\:.
\end{aligned}
\end{align}
Here $\mathcal{R}$ is the spin-independent weight function
\begin{align}
	\mathcal{R}=\prod_{j=1}^3\gamma^{(2)}(2i\alpha_j+2i\omega_1\Tilde{\alpha}_j;-i\omega_{1}r,-i\omega)\gamma^{(2)}(2i\alpha_j-i\omega_{2}(r+2\Tilde{\alpha}_j);-i\omega_{2}r,-i\omega)\; . \label{R}
\end{align}
By applying an appropriate normalization, one can embed it into the Boltzmann weight.

In terms of hyperbolic gamma functions the star-triangle relation can be written as the following integral identity 
\begin{align}
\begin{aligned}
\label{SU2identity}
	 \sum_{y=0}^{[ r/2 ]}\epsilon (y) \int _{-\infty}^{\infty} 
& \frac{\prod_{i=1}^6\gamma^{(2)}(-i(a_i\pm z)-i\omega_1(u_i\pm y);-i\omega_1r,-i\omega)}{\gamma^{(2)}(\pm 2iz\pm 2i\omega_1y;-i\omega_1r,-i\omega)}  \\
	\times & \frac{\gamma^{(2)}(-i(a_i\pm z)-i\omega_2(r-(u_i\pm y));-i\omega_2r,-i\omega)}{\gamma^{(2)}(\pm2iz-i\omega_2(r\pm2y);-i\omega_2r,-i\omega)} 
	\frac{dz}{2r\sqrt{-\omega_1\omega_2}}
    \\ =\prod_{1\leq i<j\leq 6} & \gamma^{(2)}(-i(a_i + a_j)-i\omega_1(u_i + u_j);-i\omega_1r,-i\omega)  \\ & \times 
	\gamma^{(2)}(-i(a_i + a_j)-i\omega_2(r-(u_i + u_j));-i\omega_2r,-i\omega)\; ,
\end{aligned}
\end{align}
where $\omega=\omega_1+\omega_2$ and the balancing conditions are
\begin{align}
    \nonumber \sum_{i=1}^6a_i=\omega \;, \:\:\:\:\:\:\:\:\:\:\:\:
    \nonumber \sum_{i=1}^6u_i=0\;.
\end{align}
This expression is a hyperbolic limit of lens elliptic beta integral identity \cite{Kels:2015bda, Gahramanov:2016ilb}. Note that this integral identity can also be interpreted as the equality of lens partition functions corresponding to certain supersymmetric dual theories\footnote{For further discussion of the connection between integrable spin lattice models and supersymmetric gauge theories, we refer the reader to \cite{Gahramanov:2017ysd,Yamazaki:2018xbx,Gahramanov:2022qge,Mullahasanoglu:2024gib,Mullahasanoglu:2025ont}.} \cite{Gahramanov:2016ilb}.

To construct the Sklyanin algebra in this framework, it is necessary to introduce a set of appropriate operators for the model\footnote{Here we follow the approach of \cite{Chicherin:2015mfv}.}. The integral kernel \( M(t,p) \)
\begin{align}
    M(t,p)_{z,m;x,j} = \sum_{j=0}^{r/2} \int_{-\infty}^{\infty} \frac{\gamma_h(-t+z\pm x,m-p\pm j) \gamma_h{(-t-z\pm x,-m-p\pm j)} \epsilon(j)dx}{2r \sqrt{-\omega_1\omega_2}\gamma_h(-2t,-2p)\gamma_h(\pm 2x,\pm 2j)}\;, \label{eq:Moperator}
\end{align}
satisfies a family of difference equations involving shift operators and sinusoidal factors\footnote{Such integral transforms arise naturally in the 
study of quantum integrable systems and are closely connected to the modular double of 
\( U_q(\mathfrak{sl}(2,\mathbb{R})) \) developed by Faddeev~\cite{Faddeev:1999fe}. They also appear in the explicit constructions of elliptic beta integrals and their hyperbolic limits \cite{Gahramanov:2022jxz}.}
\begin{equation}
    \begin{aligned}
         M\left(t+\frac{\omega_1}{2},p-\frac{1}{2}\right)=\frac{-2i}{\sin{\frac{2\pi}{\omega_2 r}(z+\omega_2(r-m))}}\sin{\left(\frac{\omega_1}{2}\partial_z-\frac{1}{2}\partial_{\omega_1m}\right)} M(t,p) \;, \\
      M\left(t+\frac{\omega_2}{2},p+\frac{1}{2}\right)=  \frac{2i}{\sin{\frac{2\pi}{\omega_1 r}\left(z+\omega_1 m\right)}}\sin{\left(\frac{\omega_2}{2}\partial_z+\frac{1}{2}\partial_{\omega_2 m}\right) }M(t,p) \;.
    \end{aligned}
\end{equation}
The equations above encode the functional relations obeyed by \( M(t,p) \) under discrete shifts of the variables \( t \) and \( p \). These relations can be interpreted as quantum difference equations of modular-double type, where the combined action of shift and difference operators reflects the underlying dual symmetry \( \omega_1 \leftrightarrow \omega_2 \) in the continuous parameters and \( 1 \leftrightarrow -1 \) in the discrete parameters
\begin{equation}
     \begin{aligned}
        \label{4.9}M\left(\frac{\omega_1}{2}n+\frac{\omega_2}{2}m, -\frac{n}{2}+\frac{m}{2}\right)=\left[\frac{-2i}{\sin{\frac{2\pi}{\omega_2 r}(z+\omega_2(r-m))}}\sin{\left(\frac{\omega_1}{2}\partial_z-\frac{1}{2}\partial_{\omega_1m}\right)} \right]^{n} \\ \times \left[\frac{2i}{\sin{\frac{2\pi}{\omega_1 r}(z+\omega_1 m)}}\sin{\left(\frac{\omega_2}{2}\partial_z+\frac{1}{2}\partial_{\omega_2m}\right)}\right]^{m}.
    \end{aligned}
\end{equation}
Expression (\ref{4.9}) evaluates the kernel \( M(t,p) \) at special arguments given by linear combinations of \( \omega_1 \) and \( \omega_2 \), obtained through the action of the previously derived difference operators. The integers \( n \) and \( m \) specify the number of iterations in each direction, thereby encoding the modular double symmetry via  discrete translations and trigonometric shifts.


Note that the last expression~(\ref{4.9}) requires \( M(0,0) = 1 \), 
which must be verified using residue calculus. 
The operator \( M(t,p)_{z,m;x,j} \) attains the value \((t=0, p=0)\) 
only when the corresponding limiting procedure is satisfied. 
An explicit derivation of this limit can be found in~\cite{Spiridonov:2016uae}
\begin{equation}
    \lim_{t \to 0} M(t,0)_{z,m;x,j} f(x,j) = \frac{1}{2}(f(z,m)+f(-z,r-m)) \;.
\end{equation}
This requires the \((z,m) \leftrightarrow (-z, r-m)\) symmetry in the test functions of the kernel. While the symmetry with respect to the continuous variable \( z \) is straightforward to verify, the discrete part \( m \leftrightarrow r-m \) also holds, thereby confirming that \( M(0,0) = 1 \), as required.

\section{Lens Hyperbolic Modular Double}
Now we present the algebra related to the integrable model introduced in the previous section. It satisfies the commutation relations below

\begin{equation}
    \begin{aligned}
        CA = e^{\frac{\pi}{\omega_2 r}(\omega_1+\omega_2)}AC \;, && DC = e^{\frac{\pi}{\omega_2 r}(\omega_1+\omega_2)} CD \;, \\
        [A,D]=-2i\sin^3{\frac{\pi}{\omega_2 r}(\omega_1+\omega_2)}C^2 \;,&& [B,C]=\frac{A^2-D^2}{2i\sin{\frac{\pi}{\omega_2 r}(\omega_1+\omega_2)}} \;,\\
        AB-e^{\frac{\pi}{\omega_2 r}(\omega_1+\omega_2)}BA = e^{\frac{\pi}{\omega_2 r}(\omega_1+\omega_2)} \;,&& DB-BD = \frac{i}{2}\sin{\frac{2\pi}{\omega_2 r}(\omega_1+\omega_2)}(CA-DC)\;.
    \end{aligned}
\end{equation}
This algebra arises as a trigonometric degeneration of the Sklyanin algebra studied in \cite{Gorsky:1993mh}. The generators \(A, B, C,\) and \(D\) are obtained from the hyperbolic limit of the Sklyanin generators. As expected, in the limit \( r \to 1 \), the algebra reduces to the Sklyanin algebra associated with the integrable model corresponding to the $\mathcal N=2$ supersymmetric gauge theory on the squashed sphere \( S_b^3 \) \cite{Chicherin:2015mfv}. Moreover, by taking \( q = e^{\frac{\pi}{\omega_2 r}(\omega_1+\omega_2)} \), one reproduces the commutation relations given in (\ref{trigosklyanin}).

The algebra has two Casimir operators
\begin{align}
       &K_0=e^{\frac{\pi}{\omega_2r}(\omega_1+\omega_2)}AD-\sin^2 \frac{\pi}{\omega_2r}(\omega_1+\omega_2) C^2 \;,\\
       &K_1=\frac{e^{-\frac{\pi}{\omega_2r}(\omega_1+\omega_2)}A^2+e^{\frac{\pi}{\omega_2r}(\omega_1+\omega_2)}D^2}{4\sin^2 \frac{\pi}{\omega_2r}(\omega_1+\omega_2)}-BC-\frac{1}{2}\cos \frac{\pi}{\omega_2r}(\omega_1+\omega_2) C^2\;.
\end{align}

They are intertwined by the integral kernel $M(t,p)$. This is also encouraged by the fact that Casimir operators are identical under $(t,p) \to (-t,-p)$
\begin{equation}
    \begin{aligned}
        M(t,p) A(t,p) = A(-t,-p)M(t,p)\;,&&
        M(t,p) B(t,p) = B(-t,-p)M(t,p)\;,\\
        M(t,p) C(t,p) = C(-t,-p)M(t,p)\;,&&
        M(t,p) D(t,p) = D(-t,-p)M(t,p)\;.\\
    \end{aligned}
\end{equation}

Motivated by the underlying dual symmetry, \(\omega_1 \leftrightarrow \omega_2\) for the continuous parameters and \(1 \leftrightarrow -1\) for the discrete parameters in both the structure and difference operator relations of \(M(t,p)\), this algebra naturally admits a dual under the same symmetry. Exchanging \(\omega_1 \leftrightarrow \omega_2\) and \(1 \leftrightarrow -1\) gives rise to a dual set of generators \(\tilde{A}, \tilde{B}, \tilde{C}, \tilde{D}\), which satisfy the same algebraic relations as the original generators
\begin{equation}
    \begin{aligned}
        \tilde{C}\tilde{A} = e^{\frac{\pi}{\omega_1 r}(\omega_1+\omega_2)}\tilde{A}\tilde{C} \;,&& \tilde{D}\tilde{C} = e^{\frac{\pi}{\omega_1 r}(\omega_1+\omega_2)} \tilde{C}\tilde{D}\;, \\
        [\tilde{A},\tilde{D}]=-2i\sin^3{\frac{\pi}{\omega_1 r}(\omega_1+\omega_2)}\tilde{C}^2 \;,&& [\tilde{B},\tilde{C}]=\frac{\tilde{A}^2-\tilde{D}^2}{2i\sin{\frac{\pi}{\omega_1 r}(\omega_1+\omega_2)}}\;, \\
        \tilde{A}\tilde{B}-e^{\frac{\pi}{\omega_1 r}(\omega_1+\omega_2)}\tilde{B}\tilde{A} = e^{\frac{\pi}{\omega_1 r}(\omega_1+\omega_2)} \;,&& \tilde{D}\tilde{B}-\tilde{B}\tilde{D} = \frac{i}{2}\sin{\frac{2\pi}{\omega_1 r}(\omega_1+\omega_2)}(\tilde{C}\tilde{A}-\tilde{D}\tilde{C})\;.
    \end{aligned}
\end{equation}
We would like to emphasize generators are switched to $\tilde{A}$, $\tilde{B}$, $\tilde{C}$ and $\tilde{D}$. The Casimir operators take the form
\begin{align}
&\tilde{K}_0=e^{\frac{\pi}{\omega_1r}(\omega_1+\omega_2)}\tilde{A}\tilde{D}-\sin^2 \frac{\pi}{\omega_1r}(\omega_1+\omega_2) \tilde{C}^2 \;,\\
&\tilde{K}_1=\frac{e^{-\frac{\pi}{\omega_1r}(\omega_1+\omega_2)}\tilde{A}^2+e^{\frac{\pi}{\omega_1r}(\omega_1+\omega_2)}\tilde{D}^2}{4\sin^2 \frac{\pi}{\omega_1r}(\omega_1+\omega_2)}-\tilde{B}\tilde{C}-\frac{1}{2}\cos \frac{\pi}{\omega_1r}(\omega_1+\omega_2) \tilde{C}^2\;.
\end{align}

They satisfy the exact commutation relations with the $M(t,p)$ as well. We further note that the Casimir operators remain invariant under the transformation  $(t,p) \to (-t,-p)$
\begin{equation}
    \begin{aligned}
        M(t,p) \tilde{A}(t,p) = \tilde{A}(-t,-p)M(t,p)\;,&&
        M(t,p) \tilde{B}(t,p) = \tilde{B}(-t,-p)M(t,p)\;,\\
        M(t,p) \tilde{C}(t,p) = \tilde{C}(-t,-p)M(t,p)\;,&&
        M(t,p) \tilde{D}(t,p) = \tilde{D}(-t,-p)M(t,p)\;.\\
    \end{aligned}
\end{equation}
Together, \( A, B, C, D \) and \( \tilde{A}, \tilde{B}, \tilde{C}, \tilde{D} \) generate the algebra that we refer to as the lens hyperbolic modular double.

The lens hyperbolic modular double admits a realization in terms of finite-difference operators whose coefficients depend on the representation parameters \( t \) and \( p \). In both sets of generators, these operators act on functions \( f(z,m) \) with \( z \in \mathbb{C} \) and \( m \in \mathbb{Z}_r \). In the following, we present both of them
\begin{equation}
    \begin{aligned}
       &A(t,p) = \frac{e^{\frac{\pi}{2\omega_2 r}(\omega)}e^{-\frac{i\pi}{\omega_2 r}(t-\omega_2p)}}{\sin{\frac{2\pi}{\omega_2r}(z-\omega_2m)}}\left(e^{\frac{\omega_1}{2}\partial_z-\frac{1}{2}\partial_{\omega_1m}}e^{\frac{i\pi}{\omega_2 r}(2z-2\omega_2m)}-e^{-\frac{\omega_1}{2}\partial_z+\frac{1}{2}\partial_{\omega_1m}}e^{-\frac{i\pi}{\omega_2 r}(2z-2\omega_2m)}\right) \;,\\
       & B(t,p)= -\frac{1}{2}\cos{\frac{2\pi}{\omega_2 r}(\omega)}C(t,p) \\-
       &\frac{\cos{\frac{\pi}{\omega_2 r}(4z-4\omega_2m+2t-2\omega_2p)}e^{\frac{\omega_1}{2}\partial_z-\frac{1}{2}\partial_{\omega_1m}}-\cos{\frac{\pi}{\omega_2 r}(4z-4\omega_2m-2t+2\omega_2p)}e^{-\frac{\omega_1}{2}\partial_z+\frac{1}{2}\partial_{\omega_1 m}})}{4\sin{\frac{2\pi}{\omega_2 r}(\omega)}\sin{\frac{2\pi}{\omega_2 r}(z-\omega_2 m)}}\;,\\
       &C(t,p) = \frac{1}{2\sin{\frac{\pi}{\omega_2 r}(\omega)}\sin{\frac{2\pi}{\omega_2r}(z-\omega_2m)}}\left(e^{\frac{\omega_1}{2}\partial_z-\frac{1}{2}\partial_{\omega_1m}}-e^{-\frac{\omega_1}{2}\partial_z+\frac{1}{2}\partial_{\omega_1m}}\right)\;,\\
       & D(t,p) =\frac{e^{\frac{\pi}{2\omega_2 r}(\omega)}e^{\frac{i\pi}{\omega_2 r}(t-\omega_2p)}}{\sin{\frac{2\pi}{\omega_2r}(z-\omega_2m)}}\left(e^{\frac{\omega_1}{2}\partial_z-\frac{1}{2}\partial_{\omega_1m}}e^{-\frac{i\pi}{\omega_2 r}(2z-2\omega_2m)}-e^{\frac{\omega_1}{2}\partial_z-\frac{1}{2}\partial_{\omega_1m}}e^{\frac{i\pi}{\omega_2 r}(2z-2\omega_2m)}\right)\;.
    \end{aligned}
\end{equation}
The operators $A,B,C,D$ written above correspond to the copy of the algebra
related to $\omega_2$. 
Using symmetry, exchanging $\omega_1 \leftrightarrow \omega_2$ for continuous parameters and $1 \leftrightarrow -1$ produces a second
dual copy with generators $\Tilde{A},\Tilde{B},\Tilde{C},\Tilde{D}$ which is related to $\omega_1$
\begin{equation}
    \begin{aligned}
       &\Tilde{A}(t,p) = \frac{e^{\frac{\pi}{2\omega_1 r}(\omega)}e^{-\frac{i\pi}{\omega_1 r}(t-\omega_1p)}}{\sin{\frac{2\pi}{\omega_1r}(z-\omega_1m)}}\left(e^{\frac{\omega_2}{2}\partial_z-\frac{1}{2}\partial_{\omega_2m}}e^{\frac{i\pi}{\omega_1 r}(2z-2\omega_1m)}-e^{-\frac{\omega_2}{2}\partial_z+\frac{1}{2}\partial_{\omega_2m}}e^{-\frac{i\pi}{\omega_1 r}(2z-2\omega_1m)}\right)\;, 
       \\
       & \Tilde{B}(t,p)= -\frac{1}{2}\cos{\frac{2\pi}{\omega_1 r}(\omega)}\Tilde{C}(t,p) 
       \\
       &-\frac{\cos{\frac{\pi}{\omega_1 r}(4z-4\omega_1m+2t-2\omega_1p)}e^{\frac{\omega_2}{2}}\partial_z-\frac{1}{2}\partial_{\omega_2m}-\cos{\frac{\pi}{\omega_1 r}(4z-4\omega_1m-2t+2\omega_1p)}e^{-\frac{\omega_2}{2}\partial_z+\frac{1}{2}\partial_{\omega_2m}})}{4\sin{\frac{2\pi}{\omega_1 r}(\omega)}\sin{\frac{2\pi}{\omega_1 r}(z-\omega_1 m)}}\;,
       \\
       &\Tilde{C}(t,p) = \frac{1}{2\sin{\frac{\pi}{\omega_1 r}(\omega)}\sin{\frac{2\pi}{\omega_1r}(z-\omega_1m)}}\left(e^{\frac{\omega_2}{2}\partial_z-\frac{1}{2}\partial_{\omega_2m}}-e^{-\frac{\omega_2}{2}\partial_z+\frac{1}{2}\partial_{\omega_2m}}\right)\;,
       \\
       & \Tilde{D}(t,p) =\frac{e^{\frac{\pi}{2\omega_1 r}(\omega)}e^{\frac{i\pi}{\omega_1 r}(t-\omega_1p)}}{\sin{\frac{2\pi}{\omega_1r}(z-\omega_1m)}} \left(e^{\frac{\omega_2}{2}\partial_z-\frac{1}{2}\partial_{\omega_2m}}e^{-\frac{i\pi}{\omega_1 r}(2z-2\omega_1m)}-e^{\frac{\omega_2}{2}\partial_z-\frac{1}{2}\partial_{\omega_2m}}e^{\frac{i\pi}{\omega_1 r}(2z-2\omega_1m)}\right) \;.
    \end{aligned}
\end{equation}

\section{Conclusions}

In this work, we explored the algebraic structure underlying an Ising-like integrable model and introduced the lens hyperbolic modular double as a natural extension of the hyperbolic modular double. In the appropriate limit, this algebra reduces to the standard hyperbolic modular double.

There exists a deep correspondence between exact results in supersymmetric gauge theories on compact manifolds and integrable lattice models in statistical mechanics. From this perspective, it would be natural and interesting to reformulate the results presented here in the language of supersymmetric theories.\footnote{See, for instance, \cite{Lodin:2017lrc, Nieri:2017vrb} for related developments arising from the supersymmetric framework.}

Furthermore, the Yang--Baxter equation is closely related to various problems in knot theory (see, e.g., \cite{Ozdemir:2023anx}), and exploring these connections in the context of the lens hyperbolic modular double remains an intriguing direction for future research.

The limiting case of the integral identity~(\ref{SU2identity}) can be employed \cite{Bozkurt:2020gyy, Pawelkiewicz:2013wga} to establish the orthogonality and completeness relations for the Clebsch--Gordan coefficients associated with the self-dual continuous series of \( U_q(\mathfrak{osp}(1|2)) \). Within this framework, the constructions developed in the present work may naturally be extended to the self-dual continuous series of \( U_q(\mathfrak{osp}(r|2)) \).

Finally, we note that Faddeev’s modular double underlies the algebraic structure of two-dimensional Liouville gravity (see, e.g., \cite{Fan:2021bwt, Mertens:2022aou}). It would therefore be of significant interest to investigate potential applications of the lens hyperbolic modular double to such models and to explore possible higher-rank generalizations.

\section*{Acknowledgement}
Ilmar Gahramanov would like to thank Mustafa Mullahasanoglu and Erdal Çatak for stimulating discussions on the topic of this paper. All authors are supported by the Istanbul Integrability and Stringy Topics Initiative (\href{https://istringy.org/}{istringy.org}). The work of  Ilmar Gahramanov, Ali Mert Yetkin and Reyhan Yumuşak is supported by TÜBİTAK under grant number 123F436, and Reyhan Yumuşak is also supported by TÜBİTAK under the grant number 2209-A. Ilmar Gahramanov would like to thank the Nesin Mathematics Village (Şirince, Türkiye) for its hospitality, where part of this work was carried out.

\appendix
\section{Jacobi Theta Functions}
\label{appendix1}
The definitions of the standard Jacobi theta function  $\theta_{1}(z) \equiv \theta_{1}(z|\tau)$ is given below
\begin{equation}
\theta_{1}(z|\tau) = -\sum_{n\in\mathbb{Z}} 
e^{\pi i (n+\tfrac{1}{2})^2 \tau}\,
e^{2\pi i (n+\tfrac{1}{2})(z+\tfrac{1}{2})}\;.
\end{equation}

The other three theta functions can be obtained by shifts of the argument of $\theta_1$
\begin{equation}
\theta_2(z|\tau) = \theta_1\!\left(z+\tfrac{1}{2}\,\middle|\,\tau\right)\;, 
\qquad
\theta_3(z|\tau) = e^{\frac{\pi i\tau}{4} + \pi i z}
\,\theta_2\!\left(z+\tfrac{\tau}{2}\,\middle|\,\tau\right)\;,
\qquad
\theta_4(z|\tau) = \theta_3\!\left(z+\tfrac{1}{2}\,\middle|\,\tau\right)\;.
\end{equation}

\section{Reductions of R-operator}
In this section, we aim to demonstrate that the fundamental representation of the Lax operator naturally comprises the degenerate generators on one side of the lens hyperbolic modular double. This construction will be obtained from the reduced R-operator, for which we include the derivation as well. Following this, we move to discuss the characterization of relevant representations.

We are interested in the irreducible finite-dimensional representations of the lens hyperbolic modular double at the following representation labels
\begin{equation}
    \begin{aligned}
        t_{n,m} = \frac{\omega_1}{2}(n+1)+\frac{\omega_2}{2}(m+1)\;,&&
        p_{n,m} = -\frac{n+1}{2}+ \frac{m+1}{2} \;,&& n,m\in \mathbb{Z}_{\geq 0}\;.
    \end{aligned}
\end{equation}

The function $\gamma_h(-t\pm z\pm x,-p\pm m \pm j)$ will admit all the basis vectors of this representation, where $x$ and $j$ are auxiliary parameters. As they are defined from elliptic gamma functions, it can easily be seen that our definition of the hyperbolic gamma satisfies the following linear difference equations of first order
\begin{subequations}
\begin{equation}
    \gamma_{h}(u+\omega_1,y-1;\omega_1,\omega_2)= 2i\sin\frac{\pi}{\omega_2 r}(u-\omega_2y)\gamma_{h}(u,y;\omega_1,\omega_2)\;,
\end{equation}
\begin{equation}
    \gamma_{h}(u+\omega_2,y+1;\omega_1,\omega_2)= 2i\sin\frac{\pi}{\omega_1 r}(u+\omega_1y)\gamma_{h}(u,y;\omega_1,\omega_2)\;.
\end{equation}
\end{subequations}

From the above equations, generating function $\gamma_h(-t \pm z \pm x, -p \pm m \pm j)$ has the below expansion of finite product of trigonometric functions
\begin{equation}
    \begin{aligned}
        &\gamma_h(-t+z\pm x,m-p\pm j)\gamma_h(-t-z\pm x,-m-p\pm j)= \\
         &\prod_{r=0}^{l-1}2\sin{\frac{\pi}{\omega_2 r}\left(\frac{\omega_1}{2}(l-1-2r)+\frac{\omega_2}{2}(q+1)+z\pm x+\omega_2\left(r+\frac{(l-1+2r)}{2}-\frac{(q+1)}{2}-m\mp j\right)\right)} \\
        &\prod_{t=0}^{q-1}2\sin{\frac{\pi}{\omega_1 r}\left(-\frac{\omega_1}{2}(l-1)+\frac{\omega_2}{2}(q-1-2t)+z\pm x+\omega_1\left(-\frac{(l+1)}{2}+\frac{(q-1-2t)}{2}+m\pm j\right)\right)} \\
        & \times 16\sin{\frac{\pi}{\omega_1+\omega_2}\left(\frac{\omega_1}{2}(l+1)+\frac{\omega_2}{2}(q+1)-z\pm x+\omega_1\left(-\frac{(l+1)}{2}+\frac{(q+1)}{2}-m\pm j\right)\right)}\\
        &\times \sin{\frac{\pi}{\omega_1+\omega_2}\left(\frac{\omega_1}{2}(l+1)+\frac{\omega_2}{2}(q+1)-z\pm x+\omega_2\left(r+\frac{(l+1)}{2}-\frac{(q+1)}{2}+m\mp j\right)\right)}\;.
    \end{aligned}
\end{equation}
We can then deduce the natural basis of finite-dimensional representation is the following
\begin{equation}
    \cos{\left(\frac{2\pi}{\omega_2 r}j(z-w_2m)\right)}\cos{\left(\frac{2\pi}{w_1 r}l(z+\omega_1 m)\right)}\;,
\end{equation}
where $j = \overline{0,n}$ and $l = \overline{0,m}$.

To proceed, we need to show that the integral operator solution of the Yang--Baxter equation allows us to obtain all finite-dimensional solutions of the Yang--Baxter equation as well. 

Firstly, we apply $\mathbb{R}_{12}((u_1,u_2;k_1,k_2)|(t_1,t_2;p_1,p_2))$ to $\gamma_h(-(u_1-u_2)\pm z_1 \pm m_1,-(k_1-k_2)\pm z_3 \pm m_3)\Phi(z_2,m_2)$. Here $z_3$ and $m_3$ are auxiliary parameters and $\Phi(z_2,m_2)$ is a function from the second space. Then the integral operator $\mathbb{R}_{12}$ should transform spins $\sigma_1=(z_1,m_1) \to \chi_1 = (x_1,j_1)$ and $\sigma_2=(z_2,m_2) \to \chi_2 = (x_2,j_2)$ if the function it is applied is seen as function of $\sigma_1$ and $\sigma_2$. We recall the R-operator for our integrable model to see this
\begin{equation}
    \begin{aligned}
        &\mathbb{R}_{12}((u_1,u_2;n_1,n_2)|(v_1,v_2;m_1,m_2)) \Phi(\sigma_1,\sigma_2)  \\ =
        &\sum_{j_1,j_2} \int \int S(\sigma_1')\:S(\sigma_2')\:W_{u_1-v_2,n_1-m_2}(z_1,z_2;m_1,m_2)\:W_{u_1-v_1,n_1-m_1}(z_1,x_2;m_1,j_2)\\ \times
        &W_{u_2-v_2,n_2-m_2}(x_1,z_2;j_1,m_2)\:W_{u_2-v_1,n_2-m_1}(x_1,x_2;j_1,j_2)\: \Phi((x_1,j_1),(x_2,j_2))dx_1 dx_2\;.
    \end{aligned}
\end{equation}
Then one star-triangle equation (\ref{str}) is replaced seeing $\chi_1=(x_1,j_1)$ as the common variable. This should result in $W_{u_2-t_2,k_2-p_2}(z_2,m_2;x_2,j_2)$ term alongside other Boltzmann weights as well as integral over $x_2$ and sum over $j_2$ being left. One can get rid of them by the definition of $M_{2,2}$, which is the second space version of our integral kernel (\ref{eq:Moperator})
\begin{equation}
    M_{2,2}(t,p) = \frac{1}{2r\sqrt{-w_1w_2}\gamma_h(-2t,-2p)} \sum_{j_2=0}^{r/2} \int_{-\infty}^{\infty} W_{t,p}(z_2,x_2;m_2,j_2)S(x_2,j_2) d x_2\;.
\end{equation}

Finally, we collect terms on RHS of the $M_{2,2}$ operator under an operator parentheses such that they are functions of $\sigma_2=(z_2,m_2)$ instead of $\chi_2=(x_2,j_2)$, like the integral kernel has not replaced them yet. These should reveal the following equation after representation label adjustments $(t_1 := t_{n,m},t_2:=t)$ and $(p_1:=p_{n,m}, p_2:=p)$, reduced R-operator
\small
\begin{equation}
    \begin{aligned}
     \mathbb{R}_{12}((u,n_u)|(t_{n,m},t),(p_{n,m},p))[\gamma_h(-t_{n,m}\pm z_1\pm z_3,\pm m_1-p_{n,m}\pm m_3)\Phi(z_2,m_2)] \\
        = \#. \gamma_h\left(\frac{u-t_{n,m}-t}{2}\pm z_2\pm z_3,\frac{n_u-p_{n,m}-p}{2}\pm m_2\pm m_3\right)
       \\ \times \gamma_h\left(\frac{-\eta-u-t_{n,m}-t}{2}\pm z_1\pm z_2,\frac{-n_\eta-n_u-p_{n,m}-p}{2}\pm m_1\pm m_2\right)\\ \times
        M_{2,2}\left(\frac{\omega_1 n}{2}+\frac{\omega_2 m}{2},-\frac{n}{2}+\frac{m}{2}\right)_{z_2,m_2;x_2,j_2} \bigg[ \gamma_h\left(\frac{t-t_{n,m}+t}{2}\pm z_1 \pm z_2, \frac{n_u-p_{n,m}+p}{2}\pm m_1\pm m_2\right)\\ \times 
        \gamma_h\left(\frac{-\eta-u-t_{n,m}+t}{2}\pm z_2\pm z_3,\frac{-n_\eta-n_u-p_{n,m}+p}{2}\pm m_2\pm m_3) \Phi(z_2,m_2)\right)\bigg] \;,\label{reducedR}
    \end{aligned}
\end{equation}
\normalsize
where 
\begin{equation}
    \begin{aligned}
        \#=\mathcal{R} \gamma_h(-2(\eta-(u_2-v_2)),-2(n_\eta-(n_1-m_2)))2r\sqrt{-\omega_1\omega_2}\;.
    \end{aligned}
\end{equation}
Next, we investigate this equation at fundamental representation which is simply $(t=\frac{\omega_1}{2},p=-\frac{1}{2})$. Generating function reduces to the following
\begin{equation}
    \begin{aligned}
    \gamma_h\left(-\frac{\omega_1}{2}\pm z_1 \pm z_3,\frac{1}{2}\pm m_1\pm m_3\right)  = 2\cos{\frac{2\pi}{\omega_2 r}(z_1-w_2m_1)+2\cos{\frac{2\pi}{\omega_2 r}(z_3-\omega_2m_3)}} \\ = \bold{e}_1 \cos{\frac{2\pi}{\omega_2 r}(z_3-\omega_2m_3)} +\bold{e}_2\;.
    \end{aligned}
\end{equation}
And the integral kernel $M_{2,2}(t,p)$ reduces to 
\begin{equation}
    M_{2,2}\left(\frac{\omega_1}{2},-\frac{1}{2}\right) = \frac{-1}{\sin{\frac{2\pi}{\omega_2 r}(z-\omega_2m)}}(e^{\frac{\omega_1}{2}\partial_{z}-\frac{1}{2}\partial\omega_1m}-e^{-\frac{\omega_1}{2}\partial_z+\frac{1}{2}\partial\omega_2m})\;.
\end{equation}

Substituting these into the (\ref{reducedR}), one can find the RHS of the equation is transformed to the following
\begin{equation}
    \begin{aligned}
        \left[2\cos{\frac{2\pi}{\omega_2 r}(z_1-\omega_2 m_1)}-2\cos{\frac{\pi}{\omega_2 r}(2z_2+u+t-\omega_2(2m_2+n+p))} \right]\times \\
        e^{\frac{\omega_1}{2}\partial_{z_2}-\frac{1}{2}\partial_{\omega_1m_2}}\left[2\cos{\frac{2\pi}{\omega_2 r}(z_3-\omega_2m_3)}-2\cos{\frac{2\pi}{\omega_2 r}(2z_2-u+t-\omega_2(2m_2-n+p))}\right] \\
        -\left[2\cos{\frac{2\pi}{\omega_2 r}(z_1-\omega_2m_1)}-2\cos{\frac{\pi}{\omega_2 r}(2z_2-2u_2-\omega_2(2z_2-2u_2-\omega_2(2m_2-2n_1)))}\right] \times \\
        e^{-\frac{\omega_1}{2}\partial_{z_2}+\frac{1}{2}\partial_{\omega_1m_2}}\left[2\cos{\frac{2\pi}{\omega_2 r}(z_3-\omega_2 m_3)}-2\cos{\frac{\pi}{\omega_2 r}(2z_2+u-t-\omega_2(2m_2+n-p))} \right]\;.
    \end{aligned}
\end{equation}

This allows us to write the reduced R-operator $\mathbb{R}_{1,2}(u+\frac{\omega_1}{2},n-\frac{1}{2}|(t,p)) = \mathbf{L}((u,n)|(t,p))$ in a matrix form, making use of the bases $\mathbf{e_1}$ and $\mathbf{e_2}$. In this case, the Lax matrix is the following and the degenerate generators of the algebra arise naturally as a striking result
\begin{multline}
\mathbf{L}((u,n)|(t,p)) = 2\left(
  \begin{matrix}
    -e^{-\frac{i\pi}{\omega_2 r}(u-\omega_2 m)}A(t,p) - e^{\frac{i\pi}{\omega_2 r}(u-\omega_2 m)} D(t,p) \\ \sin{\frac{\pi}{\omega_2r}(\omega_1+\omega_2)}C(t,p) 
  \end{matrix}\right.                
\\
  \left.
  \begin{matrix}
    4\sin{\frac{\pi}{\omega_2 r}(\omega_1+\omega_2)}B(t,p)-2\sin{\frac{\pi}{\omega_2 r}(\omega_1+\omega_2)}(\cos{\frac{\pi}{\omega_2 r}(2u-2\omega_2n)}+\cos{\frac{\pi}{\omega_2 r}(\omega_1+\omega_2)})C(t,p) \\
    e^{\frac{i\pi}{\omega_2 r}(u-\omega_2 m)}A(t,p)+ e^{-\frac{i\pi}{\omega_2 r}(u-\omega_2 m)} D(t,p)
  \end{matrix}\right) \;.
\end{multline}

\section{Lax Matrix}

We present the general form of the Lax matrix whose entries are difference operators involving both continuous and discrete shifts, arranged in a $2 \times 2$ matrix structure
\small
\begin{equation}
\begin{aligned}
    &\bold{L}((u,n)|(t,p))_{1,1} =\frac{1}{\sin{\frac{2\pi}{\omega_2r}(z-\omega_2m)}} \bigg(2\cos{\frac{\pi}{\omega_2 r}(2z_2-2u_2-\omega_2(2m_2-2n_2))}e^{-\frac{\omega_1}{2}\partial_{z_2}+\frac{1}{2}\partial_{(\omega_1+\omega_2)m_2}}\\
    &-2\cos{\frac{\pi}{\omega_2 r}(2z_2+2u_1-\omega_2(2m_2-2n_1))}e^{\frac{\omega_1}{2}\partial_{z_2}-\frac{1}{2}\partial_{(\omega_1+\omega_2)m_2}}\bigg) \;,\\
    &\bold{L}((u,n)|(t,p))_{1,2} =\frac{1}{\sin{\frac{2\pi}{\omega_2r}(z-\omega_2m)}} \\ 
    &\bigg(4\cos{\frac{\pi}{\omega_2 r}(2z_2+2u_1-\omega_2(2m_2+2n_1))}e^{\frac{\omega_1}{2}\partial_{z_2}-\frac{1}{2}\partial(\omega_1+\omega_2)m_2}\cos{\frac{\pi}{\omega_2 r}(2z_2-2u_2-\omega_2(2m_2-2n_2))}-\\
    &-4\cos{\frac{\pi}{\omega_2 r}(2z_2-2u_1-\omega_2(2m_2-2n_1))}e^{\frac{\omega_1}{2}\partial_{z_2}-\frac{1}{2}\partial(\omega_1+\omega_2)m_2}\cos{\frac{\pi}{\omega_2 r}(2z_2+2u_2-\omega_2(2m_2+2n_2))}\bigg)\;,
    \\ 
    &\bold{L}((u,n)|(t,p))_{2,1} =\frac{1}{\sin{\frac{2\pi}{\omega_2r}(z-\omega_2m)}}
    \big(e^{\frac{\omega_1}{2}\partial_{z_2}-\frac{1}{2}\partial(\omega_1+\omega_2)m_2}-e^{-\frac{\omega_1}{2}\partial_{z_2}+\frac{1}{2}\partial(\omega_1+\omega_2)m_2}\big)\;,\\
    &\bold{L}((u,n)|(t,p))_{2,2} =\frac{1}{\sin{\frac{2\pi}{\omega_2r}(z-\omega_2m)}} \bigg(2e^{-\frac{\omega_1}{2}\partial_{z_2}+\frac{1}{2}\partial(\omega_1+\omega_2)m_2}\cos{\frac{\pi}{\omega_2 r}(2z_2+2u_2-\omega_2(2m_2+2n_2))}
    \\
    &-2e^{\frac{\omega_1}{2}\partial_{z_2}-\frac{1}{2}\partial(\omega_1+\omega_2)m_2}\cos{\frac{\pi}{\omega_2 r}(2z_2-2u_2-\omega_2(2u_2-2n_2))}\bigg) \;.
\end{aligned}
\end{equation}
\normalsize

    \nocite{*}
 	\bibliographystyle{JHEP} 
	\bibliography{references}
    
\end{document}